\documentclass[preprint,3p,12pt]{elsarticle}

\usepackage{amsmath,scalefnt}
\usepackage{color}
\graphicspath{ {figs/} }

\makeatletter
\def\verbatim{\scalefont{0.8}\@verbatim \frenchspacing\@vobeyspaces \@xverbatim}
\makeatother

\begin{document}

\title{\vskip-3cm{\baselineskip14pt
    \begin{flushleft}
      \normalsize TTP17-011
  \end{flushleft}}
  \vskip1.5cm
  Version 3 of {\tt RunDec} and {\tt CRunDec}
}

\author{
  Florian Herren
  and
  Matthias Steinhauser
  \\[1em]
  {\small\it Institut f{\"u}r Theoretische Teilchenphysik}\\
  {\small\it Karlsruhe Institute of Technology (KIT)}\\
  {\small\it 76128 Karlsruhe, Germany}
}

\date{}

\begin{abstract}
  We present new versions of the packages {\tt RunDec} and {\tt CRunDec}
  which can be used for the running and decoupling of the strong coupling
  constant and quark masses.  Furthermore several conversion formulae for heavy
  quark masses are implemented. The new versions include five-loop
  corrections of the QCD beta function and four-loop decoupling effects.
  Furthermore, various relations between the heavy quark mass defined in the
  $\overline{\rm MS}$ and other short-distance schemes are implemented to
  next-to-next-to-next-to-leading order.
  We discuss in detail the improvements and provide several examples
  which show how {\tt RunDec} and {\tt CRunDec} can be used in
  frequently occurring situations.

  \medskip

\end{abstract}

\maketitle


\newpage


\section*{Program summary}

\begin{itemize}

\item[]{\it Title of program:}
  {\tt RunDec}, {\tt CRunDec}

\item[]{\it Available from:}
  {\tt
  https://www.ttp.kit.edu/preprints/2017/ttp17-011/
  }

\item[]{\it Computer for which the program is designed and others on which it
    is operable:}
  Any computer where {\tt Mathematica} or a {\tt C++} compiler is available.

\item[]{\it Operating system or monitor under which the program has been
    tested:} 
  Linux

\item[]{\it No. of bytes in distributed program including test data etc.:}
  $400\,000$

\item[]{\it Distribution format:} 
  source code

\item[]{\it Keywords:} 
  Quantum Chromodynamics, running coupling constant,
  running quark mass, on-shell mass, $\overline{\rm MS}$ mass, threshold masses,
  decoupling of heavy particles

\item[]{\it Nature of physical problem:} The value for the coupling constant
  of Quantum Chromodynamics, $\alpha_s^{(n_f)}(\mu)$, depends on the
  considered energy scale, $\mu$, and the number of active quark flavours,
  $n_f$. The same applies to light quark masses, $m_q^{(n_f)}(\mu)$, if they
  are, e.g., evaluated in the $\overline{\rm MS}$ scheme.  In the programs
  {\tt RunDec} and {\tt CRunDec} all relevant formulae are collected and
  various procedures are provided which allow for a convenient evaluation of
  $\alpha_s^{(n_f)}(\mu)$ and $m_q^{(n_f)}(\mu)$ using the state-of-the-art
  correction terms.  Furthermore, the programs contain several conversion
  formulae which allow to transfrom the $\overline{\rm MS}$ value of a heavy
  quark into other short-distance values or the on-shell definition.

\item[]{\it Method of solution:}
  {\tt CRunDec} is implemented in {\tt C++}. For the solution of the 
  differential equations an adaptive Runge-Kutta procedure has been
  implemented. The {\tt Mathematica} version {\tt RunDec}
  uses function provided by {\tt Mathematica} to solve the differential equations.
 
\item[]{\it Restrictions on the complexity of the problem:}
  It could be that for an unphysical choice of the input parameters the 
  results are nonsensical.

\item[]{\it Typical running time:}
  In general the run time for the individual operations is below a
  millisecond. In some cases it can increase to the order of a second.

\end{itemize}


\newpage


\section{Introduction}

The fundamental parameters of QCD are the quark masses ($m_q$) and the strong
coupling constant ($\alpha_s$) which are usually defined in the $\overline{\rm
  MS}$ scheme.  As a consequence their numerical values depend on the
renormalization scale $\mu$, which in general is of the same order as the energy scale of the
considered process, and the number of active quark flavours $n_f$. In
  practice, $\alpha_s$ and $m_q$ are determined for certain values of $\mu$
  and $n_f$. Afterwards they enter predictions which require other choices. It is thus
crucial to have at hand precise relations for $m_q$ and $\alpha_s$ evaluated
at different renormalization scales and different number of active flavours.

For the heavy quark masses there are two widely used renormalization schemes,
the on-shell and the $\overline{\rm MS}$ scheme.  Depending upon the
considered physical process, there are further short-distance definitions
which have advantageous properties and are thus often used to parametrize
perturbative predictions.  It is important to have available precise relations
between the various schemes in order not to lose precision due to the
transformations.

The {\tt Mathematica} program {\tt RunDec}~\cite{Chetyrkin:2000yt} and its
{\tt C++} counterpart {\tt CRunDec}~\cite{Schmidt:2012az} achieve these
requirements: they implement the highest available perturbative QCD corrections
both for the running and decoupling of $\alpha_s$ and $m_q$, and for the
conversion of the heavy quark masses among different schemes.  In this paper we
describe versions~3 of {\tt RunDec} and {\tt CRunDec} which contain several
improvements. They are described in Section~\ref{sec::news}.
Section~\ref{sec::examples} contains several practical examples. In this paper
we refrain from presenting explicit formulae which are implemented in the
routines.  They can either be found in Ref.~\cite{Chetyrkin:2000yt}, in the
original papers where the higher order corrections have been computed, or can
easily be extracted from the source codes of {\tt RunDec} and {\tt CRunDec}.

The use of the {\tt Mathematica} version {\tt RunDec} is straightforward.
After loading {\tt RunDec.m} all {\tt Mathematica}-modules are available 
and can
immediately be used. {\tt CRunDec} is written in {\tt C++} which offers
several possibilities to access the implemented functions. In the following
we present two skeleton files which exemplify the usage. They can easily be
adapted to the problem at hand. On the one hand it is possible to work with
pointers to an object of the type {\tt CRunDec} and access the member
functions correspondingly:

\begin{verbatim}
#include <iostream>
#include "CRunDec.h"
using namespace std;
int main(){
  CRunDec * <pointer> = new CRunDec();
  double <result> = <pointer> -> <function>(<parameters>);
  return(0);
}
\end{verbatim}
Alternatively, also the following realization is possible:

\begin{verbatim}
#include <iostream>
#include "CRunDec.h"
using namespace std;
int main(){
  CRunDec <object>;
  double <result> = <object>.<function>(<parameters>);
  return(0);
}
\end{verbatim}
Several explicit examples are discussed in Section~\ref{sec::examples}. They are
illustrated with the help of {\tt Mathematica} code; the corresponding {\tt
  C++} code can be found in the ancillary files to this paper.
Section~\ref{sec::sum} contains a brief summary.


\section{\label{sec::news}New and updated routines in {\tt RunDec} and {\tt CRunDec}}

All routines connected to the running (of $\alpha_s$ and $m_q$) and to the
decoupling of heavy quarks have been updated. In all of them it is now
possible to use five-loop running accompanied by four-loop
decoupling.  We have also updated the routines establishing the relations
between $\alpha_s$ and the QCD scale parameter $\Lambda$ which allow for a more precise
extraction of $\Lambda$.  The corresponding analytic expressions
have been obtained in the following papers: The four-loop decoupling constants
for $\alpha_s$ is available from Refs.~\cite{Schroder:2005hy,Chetyrkin:2005ia}
and the one for the light quark masses has been computed
in~\cite{Liu:2015fxa}. The five-loop anomalous dimension of the quark masses
has been computed by two
groups~\cite{Baikov:2014qja,Luthe:2016xec,Baikov:2017ujl} and the five-loop
beta function is available
from~\cite{Baikov:2016tgj,Herzog:2017ohr,Luthe:2017ttc}.

As a second major update we have implemented the four-loop relation between
the heavy quark masses defined in the $\overline{\rm MS}$ and on-shell
scheme~\cite{Marquard:2015qpa,Marquard:2016dcn}. In this context a couple of
new routines have been introduced which allow the transformation between the
$\overline{\rm MS}$ or on-shell and so-called threshold masses. In particular,
we have implemented the potential subtracted (PS)~\cite{Beneke:1998rk}, the
1S~\cite{Hoang:1998hm,Hoang:1998ng,Hoang:1999zc} and the renormalon subtracted
(RS)~\cite{Pineda:2001zq} masses.
In Section~\ref{sec::examples} we show how the new routines can be used
to obtain the $\overline{\rm MS}$ mass from an experimentally determined
threshold mass including the uncertainty on the strong coupling constant.

We refrain from providing analytic results for the new perturbative
corrections which have been implemented in {\tt RunDec} and {\tt CRunDec}.
One can either find them in the original papers or easily extract them
from the {\tt Mathematica} source code. For example, the decoupling
constant which establishes the transition from $\alpha_s$
defined in the $n_f$-flavour theory to $\alpha_s$ in the $(n_f-1)$-flavour theory
using the $\overline{\rm MS}$ definition of the heavy quark mass,
can be found up to four loops with the help of

\begin{verbatim}
  RunDec`Modules`as6to5ms /. setdec
\end{verbatim}
The inverted relations and the ones where different renormalization schemes
for the heavy quark mass have been chosen are contained in the
variables        \verb|as6to5si|, \verb|as6to5os|,
\verb|as5to6ms|, \verb|as5to6si| and \verb|as5to6os|.
Analytic results up to three loops are given in Eqs.~(20) to~(25)
of Ref.~\cite{Chetyrkin:2000yt}.
The coefficients of the QCD $\beta$ function and the mass anomalous dimension
can be extracted from the variables \verb|setbeta| and \verb|setgamma|,
respectively.

In the following subsections we provide merely the name of the function in
case it has only been extended to higher loop order.  For a detailed
description of the arguments we refer to Refs.~\cite{Chetyrkin:2000yt}
and~\cite{Schmidt:2012az}.  For the new functions and for the functions where
additional arguments have been introduced\footnote{Note that the functions
  have been overloaded such that it is still possible to call them with the
  old argument list. The only exception is the argument handling the light
  quark masses in the relation between the $\overline{\rm MS}$ and on-shell
  masses as implemented in {\tt CRunDec}, see Subsection~\ref{sub::mass}.} we display the
complete function header.

\subsection{Running and decoupling of $\alpha_s$ and $m_q$}

The following functions related to the running of
$\alpha_s$ and $m_q$ have been updated to five-loop accuracy:
\\
\verb|LamImpl|,
\verb|LamExpl|,
\verb|AlphasLam|,
\\
\verb|AlphasExact|,
\verb|mMS2mMS|,
\verb|mMS2mSI|,
\verb|AsmMSrunexact|,
\\
\verb|mMS2mRGI|,
\verb|mMS2mRGImod|.
\\
Correspondingly we implemented the four-loop decoupling relations
into the following routines:
\\
\verb|DecAsUpOS|, \verb|DecAsDownOS|,
\verb|DecAsUpMS|, \verb|DecAsDownMS|,
\verb|DecAsUpSI|, \verb|DecAsDownSI|,
\\
\verb|DecMqUpOS|, \verb|DecMqDownOS|,
\verb|DecMqUpMS|, \verb|DecMqDownMS|,
\verb|DecMqUpSI|, \verb|DecMqDownSI|,
\\
\verb|DecLambdaUp|, \verb|DecLambdaDown|.
\\
As a consequence also the functions
\verb|AlL2AlH| and \verb|AlH2AlL|,
which perform a combined running and decoupling
can be used at five-loop order.

\subsection{\label{sub::mass}Quark mass relations}

The functions implementing the transitions between the heavy quark
masses defined in the on-shell and $\overline{\rm MS}$ scheme
have been extended to four loops. Since the non-logarithmic four-loop
term is not known analytically but has an uncertainty
of the order of $0.2\%$~\cite{Marquard:2016dcn} we have introduced an additional
argument for the following functions where this can be taken into account

\begin{verbatim}
mOS2mMS[mOS_,mq_,asmu_,mu_,nnf_,loops_,fdelm_]
mOS2mMS[mOS_,mq_,asmu_,mu_,nnf_,loops_]
mOS2mSI[mOS_,mq_,asM_,nnf_,loops_,fdelm_]
mOS2mSI[mOS_,mq_,asM_,nnf_,loops_]
mMS2mOS[mMS_,mq_,asmu_,mu_,nnf_,loops_,fdelm_]
mMS2mOS[mMS_,mq_,asmu_,mu_,nnf_,loops_]
\end{verbatim}
Here \verb|fdelm| multiplies the non-logarithmic four-loop
coefficient of order $\alpha_s^4$. The default value is 1.
A $0.2\%$ variation is obtained with \verb|fdelm|=0.998.
The function prototype of the corresponding
routines in {\tt CRunDec} read

\begin{verbatim}
double mOS2mMS(double mOS, std::pair<double,double>* mq,
               double asmu, double mu, int nf, int nloops, double fdelm);
double mOS2mMS(double mOS, std::pair<double,double>* mq,
               double asmu, double mu, int nf, int nloops);
double mOS2mSI(double mOS, std::pair<double,double>* mq,
               double asM,int nf, int nloops, double fdelm);
double mOS2mSI(double mOS, std::pair<double,double>* mq,
               double asM,int nf, int nloops);
double mMS2mOS(double mMS, std::pair<double,double>* mq,
               double asmu, double mu, int nf, int nloops, double fdelm);
double mMS2mOS(double mMS, std::pair<double,double>* mq,
               double asmu, double mu, int nf, int nloops);
\end{verbatim}
The arguments of the following functions are unchanged:
\\
\verb|mOS2mMSrun|, \verb|mOS2mMSit|, \verb|mMS2mOSrun|.

A further improvement in these functions is related to the light quark mass
effects. Up to now they have only been implemented up to two-loop
order~\cite{Gray:1990yh}.  We have extended the implementation to three loops
using the results of Ref.~\cite{Bekavac:2007tk}. The light quark masses,
$m_q$, are renormalized in the $\overline{\rm MS}$ scheme at their own
renormalization scale $\mu_q$.  In the previous versions of {\tt RunDec} the
argument \verb|mq| was a list containing the light quark masses.  For example,
finite charm and bottom quark masses in the top quark mass relation are taken
into account via \verb|{4.163,0.986}|. After including the three-loop
corrections the scale $\mu_q$ has to be specified and the argument \verb|mq|
has been extended to \verb|{{4.163,4.163},{0.986,3.0}}| to specify $m_b(m_b)$
and $m_c(3~\mbox{GeV})$. Note that it is still possible to provide a
non-nested list. In that case the quark masses are interpreted as $m_q(m_q)$.

In the {\tt C++} version modifications in the parameter passing were
necessary, which are incompatible with older versions in case non-zero light
quark masses are used. In version~3 an array of pairs of \verb|double|
variables has to be passed instead of a \verb|double| array with four
elements. The first element of each pair is the mass, the second is the scale.
Note that the array must either contain four elements or can be a null
pointer.

The four-loop relation between the $\overline{\rm MS}$ and
on-shell quark mass can be used to derive the relations to the
so-called threshold masses to the corresponding accuracy (see
Ref.~\cite{Marquard:2016dcn} for more details).
In particular, we have implemented the following relations
to the potential subtracted mass (\verb|PS|),
to the 1S mass (\verb|1S|),
to the renormalon subtracted mass (\verb|RS|),
and to a modified version of the latter (\verb|RSp|):

\begin{verbatim}
mOS2mPS[mOS_,mq_,asmu_,mu_,muf_,nl_,loops_]
mMS2mPS[mMS_,mq_,asmu_,mu_,muf_,nl_,loops_]
mMS2mPS[mMS_,mq_,asmu_,mu_,muf_,nl_,loops_,fdelm_]
mPS2mMS[mPS_,mq_,asnlmu_,mu_,muf_,nl_,loops_]
mPS2mMS[mPS_,mq_,asnlmu_,mu_,muf_,nl_,loops_,fdelm_]
mPS2mSI[mPS_,mq_,asfct_,muf_,nl_,loops_]
mPS2mSI[mPS_,mq_,asfct_,muf_,nl_,loops_,fdelm_]
\end{verbatim}
\begin{verbatim}
mOS2m1S[mOS_,mq_,asmu_,mu_,nl_,loops_]
mMS2m1S[mMS_,mq_,asmu_,mu_,nl_,loops_]
mMS2m1S[mMS_,mq_,asmu_,mu_,nl_,loops_,fdelm_]
m1S2mMS[m1S_,mq_,asnlmu_,mu_,nl_,loops_]
m1S2mMS[m1S_,mq_,asnlmu_,mu_,nl_,loops_,fdelm_]
m1S2mSI[m1S_,mq_,asfct_,nl_,loops_] 
m1S2mSI[m1S_,mq_,asfct_,nl_,loops_,fdelm_]
\end{verbatim}
\begin{verbatim}
mOS2mRS[mOS_,mq_,asmu_,mu_,nuf_,nl_,loops_]
mMS2mRS[mMS_,mq_,asmu_,mu_,nuf_,nl_,loops_]
mMS2mRS[mMS_,mq_,asmu_,mu_,nuf_,nl_,loops_,fdelm_]
mRS2mMS[mRS_,mq_,asnlmu_,mu_,nuf_,nl_,loops_]
mRS2mMS[mRS_,mq_,asnlmu_,mu_,nuf_,nl_,loops_,fdelm_]
mRS2mSI[mRS_,mq_,asfct_,nuf_,nl_,loops_]
mRS2mSI[mRS_,mq_,asfct_,nuf_,nl_,loops_,fdelm_]
\end{verbatim}
\begin{verbatim}
mOS2mRSp[mOS_,mq_,asmu_,mu_,nuf_,nl_,loops_]
mMS2mRSp[mMS_,mq_,asmu_,mu_,nuf_,nl_,loops_]
mMS2mRSp[mMS_,mq_,asmu_,mu_,nuf_,nl_,loops_,fdelm_]
mRSp2mMS[mRS_,mq_,asnlmu_,mu_,nuf_,nl_,loops_]
mRSp2mMS[mRS_,mq_,asnlmu_,mu_,nuf_,nl_,loops_,fdelm_]
mRSp2mSI[mRS_,mq_,asfct_,nuf_,nl_,loops_]
mRSp2mSI[mRS_,mq_,asfct_,nuf_,nl_,loops_,fdelm_]
\end{verbatim}
In the {\tt C++} version the function prototypes are given by

\begin{verbatim}
double mOS2mPS(double mOS, std::pair<double,double>* mq, double asmu, 
               double mu, double muf, int nl, int nloops);
double mMS2mPS(double mMS, std::pair<double,double>* mq, double asmu,
               double mu, double muf, int nl, int nloops);
double mMS2mPS(double mMS, std::pair<double,double>* mq, double asmu,
               double mu, double muf, int nl, int nloops, double fdelm);
double mPS2mMS(double mPS, std::pair<double,double>* mq, double asmu,
               double mu, double muf, int nl, int nloops);
double mPS2mMS(double mPS, std::pair<double,double>* mq, double asmu,
               double mu, double muf, int nl, int nloops, double fdelm);
double mPS2mSI(double mPS, std::pair<double,double>* mq, double (*as)(double),
               double muf, int nl, int nloops);
double mPS2mSI(double mPS, std::pair<double,double>* mq, double (*as)(double),
               double muf, int nl, int nloops, double fdelm);

double mOS2m1S(double mOS, std::pair<double,double>* mq, double asmu,
               double mu, int nl, int nloops);
double mMS2m1S(double mMS, std::pair<double,double>* mq, double asmu,
               double mu, int nl, int nloops);
double mMS2m1S(double mMS, std::pair<double,double>* mq, double asmu,
               double mu, int nl, int nloops, double fdelm);
double m1S2mMS(double m1S, std::pair<double,double>* mq, double asmu,
               double mu, int nl, int nloops);
double m1S2mMS(double m1S, std::pair<double,double>* mq, double asmu,
               double mu, int nl, int nloops, double fdelm);
double m1S2mSI(double m1S, std::pair<double,double>* mq,
               double (*as)(double), int nl, int nloops);
double m1S2mSI(double m1S, std::pair<double,double>* mq,
               double (*as)(double), int nl, int nloops, double fdelm);

double mOS2mRS(double mOS, std::pair<double,double>* mq, double asmu,
               double mu, double nuf, int nl, int nloops);
double mMS2mRS(double mMS, std::pair<double,double>* mq, double asmu,
               double mu, double nuf, int nl, int nloops);
double mMS2mRS(double mMS, std::pair<double,double>* mq, double asmu,
               double mu, double nuf, int nl, int nloops, double fdelm);
double mRS2mMS(double mRS, std::pair<double,double>* mq, double asmu,
               double mu, double nuf, int nl, int nloops);
double mRS2mMS(double mRS, std::pair<double,double>* mq, double asmu,
               double mu, double nuf, int nl, int nloops, double fdelm);
double mRS2mSI(double mRS, std::pair<double,double>* mq, double (*as)(double),
               double nuf, int nl, int nloops);
double mRS2mSI(double mRS, std::pair<double,double>* mq, double (*as)(double),
               double nuf, int nl, int nloops, double fdelm);

double mOS2mRSp(double mOS, std::pair<double,double>* mq, double asmu,
                double mu, double nuf, int nl, int nloops);
double mMS2mRSp(double mMS, std::pair<double,double>* mq, double asmu,
                double mu, double nuf, int nl, int nloops);
double mMS2mRSp(double mMS, std::pair<double,double>* mq, double asmu,
                double mu, double nuf, int nl, int nloops, double fdelm);
double mRSp2mMS(double mRS, std::pair<double,double>* mq, double asmu,
                double mu, double nuf, int nl, int nloops);
double mRSp2mMS(double mRS, std::pair<double,double>* mq, double asmu,
                double mu, double nuf, int nl, int nloops, double fdelm);
double mRSp2mSI(double mRS, std::pair<double,double>* mq, double (*as)(double),
                double nuf, int nl, int nloops);
double mRSp2mSI(double mRS, std::pair<double,double>* mq, double (*as)(double),
                double nuf, int nl, int nloops, double fdelm);
\end{verbatim}
The meaning of the arguments is in close analogy to the 
function implementing the $\overline{\rm MS}$-on-shell relation,
see also the description of the individual modules in the {\tt Mathematica}
version. In particular, also the argument \verb|mq| for light quark mass
effects is present. However, in the relations involving threshold masses it is
not active.

For the relations involving PS and RS masses additional factorization scales
are introduced which are denoted by \verb|muf| and \verb|nuf|,
respectively.\footnote{See the original publications~\cite{Beneke:1998rk}
  and~\cite{Pineda:2001zq} for more details.}
Note that for technical reasons it is necessary to provide
just the function name and not the numerical value of $\alpha_s$
for those functions which convert to the scale-invariant mass
(``\verb|SI|'').
This is denoted by \verb|asfact| in the argument list.
In the {\tt C++} version a pointer to a function taking the renormalization scale
$\mu$ as argument has to be passed instead.

In {\tt RunDec} there is also the routine {\tt
  AsRunDec} which extracts from the renormalization scales specified in the
input the necessary information about the decoupling steps. Since for some
cases this might lead to inconsistencies we recommend not to use this module
but use instead \verb|AlL2AlH| or \verb|AlH2AlL|, see also the examples in
Section~\ref{sec::examples}.  For this reason we do not maintain this
routine. This also concerns the versions of \verb|mOS2mMS| and \verb|mMS2mOS|
with the headers \verb|mOS2mMS[mOS_,nnf_,loops_]| and
\verb|mMS2mOS[mMS_,nnf_,loops_]|.


\section{\label{sec::examples}Useful examples}

In this section we present various concrete examples which 
are useful for many everyday situations. We discuss in detail
numerical effects and provide the source codes which can easily
be adapted to the own programs.

The examples require various input values which we choose as
follows~\cite{Olive:2016xmw,Chetyrkin:2009fv,Baikov:2008jh}
{\small\begin{align}
  ({\tt asMz})   && \alpha_s^{(5)}(M_Z) &= 0.1181 \pm 0.0011 &
  ({\tt asMtau}) && \alpha_s^{(5)}(m_\tau) &= 0.332 \pm 0.016
  \nonumber\\
  ({\tt Mz}) && M_Z &= 91.1876 \pm 0.0021~\mbox{GeV} &
  ({\tt Mh}) && M_H &= 125.09 \pm 0.24~\mbox{GeV}
  \nonumber\\
  ({\tt muc}) && m_c(m_c) &= 1.279 \pm  0.013~\mbox{GeV} &
  ({\tt mc3})&& m_c(3~\mbox{GeV}) &= 0.986 \pm 0.013~\mbox{GeV}
  \nonumber\\
  ({\tt mub})  && m_b(m_b) &= 4.163 \pm 0.016~\mbox{GeV} &
  ({\tt Mtau}) && m_\tau &= 1.77686 \pm 0.00012~\mbox{GeV}
  \nonumber\\
  ({\tt Mc})   && M_c &= 1.5~\mbox{GeV} &
  ({\tt Mb}) && M_b &= 4.8~\mbox{GeV}
  \nonumber\\
  ({\tt Mt}) && M_t &= 173.21 \pm 0.87~\mbox{GeV}
  \label{eq::input}
\end{align}}
The symbols inside the brackets show the notation used in {\tt RunDec} and {\tt
  CRunDec}.  The central values (without uncertainties) are implemented in the
variable \verb|NumDef| (both in {\tt RunDec} and {\tt CRunDec}).

\subsection{Running and decoupling}


\subsubsection{$\alpha_s$ at high energies}

As a typical example we consider the extraction of $\alpha_s$ from the two- to
three-jet event ratio as, for example, performed by
CMS~\cite{Chatrchyan:2013txa}.  For $\mu=896$~GeV the value
$\alpha_s(\mu)=0.0889\pm0.0034$ has been obtained.  Interpreting this result
as the strong coupling in a six-flavour theory we can use

\begin{verbatim}
  muin  = 896;
  as6mu = 0.0889;
  as5Mz = AlH2AlL[as6mu, muin, {{6, Mt /. NumDef, 2*Mt /. NumDef}}, Mz /. NumDef, 5];
\end{verbatim}
and obtain $\alpha_s^{(5)}(M_Z)=0.1170$. Performing the decoupling for
$\mu_{\rm dec}^{(t)}=M_t$ instead of $\mu_{\rm dec}^{(t)}=2M_t$
does not change the provided four
digits. The same is true if four-loop (instead of five-loop) accuracy is used
for the running.  If, however, \verb|as6mu| is interpreted in the five-flavour
theory one obtains $\alpha_s^{(5)}(M_Z)=0.1199$ after using

\begin{verbatim}
  AlphasExact[as6mu, muin, Mz /. NumDef, 5, 5]
\end{verbatim}
Note that \verb|AlH2AlL| implements the on-shell decoupling constants. In case
the $\overline{\rm MS}$ or scale-invariant (``\verb|SI|'') version shall be used
one has to combine \verb|AlphasExact| and \verb|DecAsDownMS| or
\verb|DecAsDownSI| to obtain $\alpha_s^{(5)}(M_Z)$.

As a further example we consider the recent determination of $\alpha_s$ 
from the double-differential inclusive jet cross section
as reported by the CMS collaboration in Ref.~\cite{Khachatryan:2016mlc}.
The measured transverse momentum ($p_{\mathrm{T}}$) interval is
divided into several ranges and in each of them the strong coupling constant
is extracted. In the last row of Tab.~5 of Ref.~\cite{Khachatryan:2016mlc}
one finds $\alpha_s^{(5)}(M_Z) = 0.1162_{-0.0062}^{+0.0070}$
which is obtained from an energy scale of $Q=1508.04$~GeV.
In Ref.~\cite{Khachatryan:2016mlc} two-loop running with
five active flavours has been used to obtain
$\alpha_s^{(5)}(Q) = 0.0822_{-0.0031}^{+0.0034}$.
Within {\tt RunDec} this result can be obtained via

\begin{verbatim}
asmz      = 0.1162;
asmzplus  = asmz + 0.0070;
asmzminus = asmz - 0.0062;
Q         = 1508.04;

nloops = 2;

asQ            = AlphasExact[asmz, Mz /. NumDef, Q, 5, nloops];
delasQplus     = Abs[asQ - AlphasExact[asmzplus, Mz /. NumDef, Q, 5, nloops]];
delasQminus    = Abs[asQ - AlphasExact[asmzminus, Mz /. NumDef, Q, 5, nloops]];
truncuncert    = Abs[asQ - AlphasExact[asmz, Mz /. NumDef, Q, 5, nloops-1]];
totuncertplus  = Sqrt[truncuncert^2 + delasQplus^2];
totuncertminus = Sqrt[truncuncert^2 + delasQminus^2];
\end{verbatim}
where we have introduced an additional uncertainty due to the truncation
of the perturbative series. We obtain
$\alpha_s^{(5)}(Q) = 0.0822_{-0.0032}^{+0.0035}$.
Choosing five-loop accuracy, i.e. \verb|nloops = 5|, leads to
$\alpha_s^{(5)}(Q) = 0.0822_{-0.0031}^{+0.0034}$
and thus has only a minor effect on the result.

At energy scales above a few hundred GeV also the top quark is an active quark
flavour and, in principle, one has to include decoupling effects when
relating $\alpha_s^{(5)}(M_Z)$ to $\alpha_s^{(6)}(Q)$. With the help
of {\tt RunDec} this can be taken into account as follows

\begin{verbatim}
nloops   = 5;

dec            = {{6, Mt /. NumDef, 2*Mt /. NumDef}};
asQ            = AlL2AlH[asmz, Mz /. NumDef, dec, Q, nloops];
delasQplus     = Abs[asQ - AlL2AlH[asmzplus, Mz /. NumDef, dec, Q, nloops]];
delasQminus    = Abs[asQ - AlL2AlH[asmzminus, Mz /. NumDef, dec, Q, nloops]];
truncuncert    = Abs[asQ - AlL2AlH[asmz, Mz /. NumDef, dec, Q, nloops-1]];
totuncertplus  = Sqrt[truncuncert^2 + delasQplus^2];
totuncertminus = Sqrt[truncuncert^2 + delasQminus^2];
\end{verbatim}
It is furthermore possible to vary the decoupling scale between 
$M_t/f < \mu_{\mathrm{dec}}^{(t)} < f M_t$ where $f$ is often chosen between $2$ and $4$.
Within {\tt Mathematica} this can be realized for $f=4$ via

\begin{verbatim}
f    = 4;
step = 5;
tmp  = Map[ AlL2AlH[asmz, N[Mz /. NumDef], {{6, N[Mt /. NumDef], #}}, Q, nloops]&, 
            Range[N[Mt /. NumDef]/f, f*N[Mt /. NumDef], step]];
scaleuncert = Max[ tmp ] - Min[ tmp ];
\end{verbatim}
Using five-loop accuracy we arrive at $\alpha_s^{(6)}(Q) =
0.0840_{-0.0033}^{+0.0036}$ which differs significantly from the value for
$\alpha_s^{(5)}(Q)$ given above. Here, the scale uncertainty
(``\verb|scaleuncert|'') is negligible.  For \verb|nloops = 2| we obtain
$\alpha_s^{(6)}(Q) = 0.0839_{-0.0033}^{+0.0036}$ where 10\% of the error is
due to scale uncertainty.


\subsubsection{\label{sub::tau}Compute $\alpha_s^{(5)}(M_Z)$ from $\alpha_s^{(3)}(m_\tau)$}

As a further example where higher order corrections are crucial, we consider the
extraction of $\alpha_s^{(5)}(M_Z)$ from $\tau$ lepton decays. In this
case one has to take into account the effect of two flavour thresholds when running
from $m_\tau$ to $M_Z$. Using \verb|AlphasExact| and
\verb|DecAsUpSI| we can calculate $\alpha_s^{(5)}(M_Z)$ to $n$-loop accuracy
with the help of

\begin{verbatim}
As5MZ[asin_?NumberQ, mu_?NumberQ, thr1_?NumberQ, thr2_?NumberQ, n_?IntegerQ] := Module[
    {as3c, as4c, as4b, as5b, as5, as5mz},
    as3c  = AlphasExact[asin, mu, thr1, 3, n];
    as4c  = DecAsUpSI[as3c, muc /. NumDef, thr1, 3, n];
    as4b  = AlphasExact[as4c, thr1, thr2, 4, n];
    as5b  = DecAsUpSI[as4b, mub /. NumDef, thr2, 4, n];
    as5mz = AlphasExact[as5b, thr2, Mz /. NumDef, 5, n];
    Return[as5mz];
];
\end{verbatim}
This module can be used to obtain $\alpha_s^{(5)}(M_Z)$ including the
uncertainties due to truncation of the perturbative series and the scale
variation introduced by matching the three- and four-, as well as four-
and five-flavour theory.  Following Ref.~\cite{Baikov:2008jh} we calculate the
truncation uncertainty by taking the difference between $n$-loop and
$(n-1)$-loop results for $\alpha_s^{(5)}(M_Z)$. We obtain the scale
uncertainties by varying the decoupling scale between $\mu_{\mathrm{dec}}/3$
and $3 \mu_{\mathrm{dec}}$ around $\mu^{(c)}_{\mathrm{dec}} = 3.0$ GeV for charm and
$\mu^{(b)}_{\mathrm{dec}} = m_b(m_b)$ for bottom.  Using $\alpha_s^{(3)}(m_\tau) = 0.332
\pm 0.016$~\cite{Baikov:2008jh}, the scale invariant charm and bottom quark
masses, and the following {\tt RunDec} commands

\begin{verbatim}
mudecc      = 3.0;
mudecb      = mub /. NumDef;
asmtauerror = 0.016;
as5         = As5MZ[asMtau /. NumDef, Mtau /. NumDef, mudecc, mudecb, 5];
f    = 3;
step = 1;
truncerr    = Abs[as5 - As5MZ[asMtau /. NumDef, Mtau /. NumDef, mudecc, mudecb, 5-1]];
tmp = Map[ As5MZ[asMtau /. NumDef, Mtau /. NumDef, #, mudecb, 5]&,
           Range[mudecc/f, f*mudecc, step]];
scaleuncertc = Max[ tmp ] - Min[ tmp ];
tmp = Map[ As5MZ[asMtau /. NumDef, Mtau /. NumDef, mudecc, #, 5]&, 
           Range[mudecb/f, f*mudecb, step]];
scaleuncertb = Max[ tmp ] - Min[ tmp ];
totaluncert  = Sqrt[truncerr^2 + scaleuncertc^2 + scaleuncertb^2];
expuncert    = Abs[as5 - As5MZ[(asMtau /. NumDef)-asmtauerror, Mtau /. NumDef,
                             mudecc, mudecb, 5]];
\end{verbatim}
one obtains $\alpha_s^{(5)}(M_Z) = 0.1201 \pm 0.0019 \pm 0.0003$. The
first uncertainty reflects the error of $\alpha_s^{(3)}(m_\tau)$ and the
second one the truncation and scale uncertainties which amount to
$\delta_{\mathrm{trunc}} = 0.00005$, $\delta_b = 0.00003$ and $\delta_c =
0.00030$, respectively. It is clearly dominated by the scale uncertainty due
to the decoupling of the charm quark.  In Tab.~\ref{tbl:asmtau}
$\sqrt{\delta_{\mathrm{trunc}}^2 + \delta_b^2 + \delta_c^2}$ is shown as a
function of the number of loops used for the running of $\alpha_s$.  One
observes a dramatic reduction after including three loops.  Afterwards the
improvements are significantly smaller than the uncertainty of
$\alpha_s^{(3)}(m_\tau)$.

\begin{table}[t]
  \begin{center}
    \begin{tabular}{c | c}
      $\#$ loops & $\delta = \sqrt{\delta_{\mathrm{trunc}}^2 + \delta_b^2 + \delta_c^2}$\\
      \hline
      2 & $0.0054$\\
      3 & $0.0009$\\
      4 & $0.0005$\\
      5 & $0.0003$\\
    \end{tabular}
    \caption{\label{tbl:asmtau} Uncertainty induced by truncation
      ($\delta_{\mathrm{trunc}}$) and decoupling scale uncertainties
      ($\delta_b$ and $\delta_c$) in the renormalization group running of
      $\alpha_s$ from $m_\tau$ to $M_Z$ as a function of the number of loops
      used for the beta function.}
  \end{center}
\end{table}

\begin{figure}[t]
  \centering
  \begin{tabular}{cc}
    \hspace*{-2em}
    \includegraphics[width=.49\textwidth]{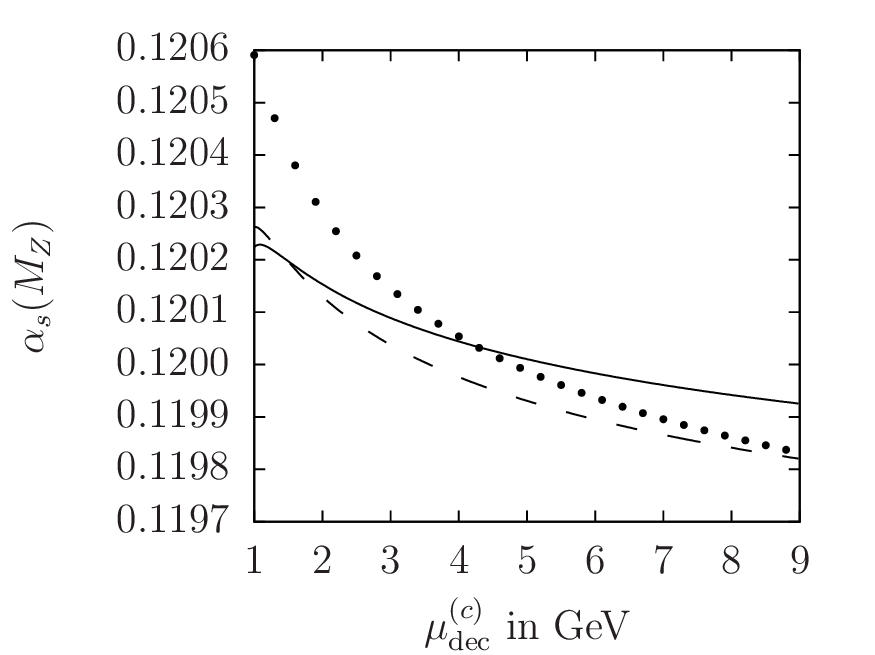} &
    \hspace*{-2em}
    \includegraphics[width=.49\textwidth]{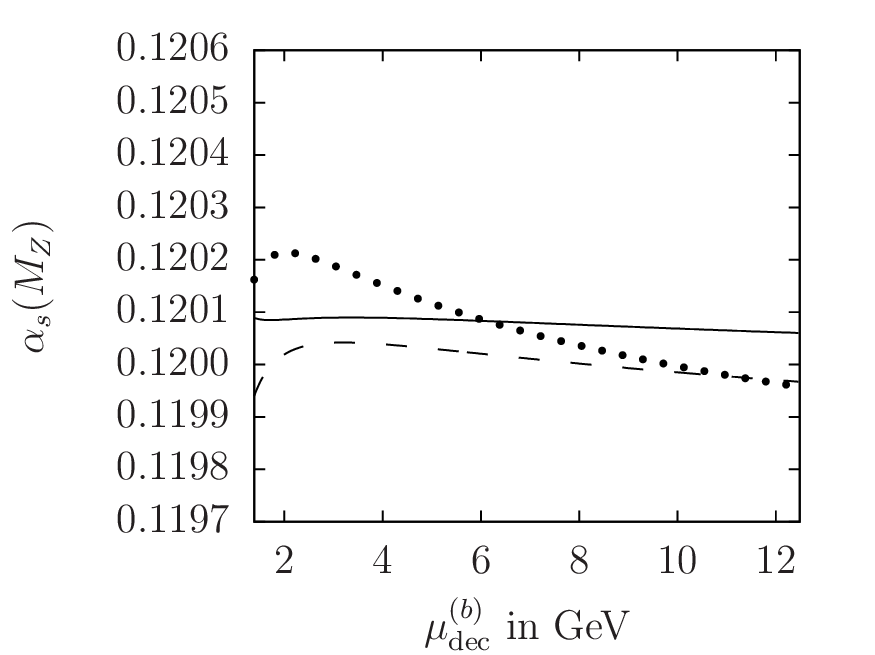}
  \end{tabular}
  \caption{\label{fig::mudecbc}
    $\alpha^{(5)}_s(M_Z)$ calculated from $\alpha^{(3)}_s(m_\tau)$ with the
    charm (left) and bottom (right) quark decoupling scale varied 
    between 1 GeV and 9 GeV for charm and $m_b/3\approx 1.4$~GeV and
    $3m_b\approx12.6$~GeV for bottom. The dotted, dashed and solid curves
    correspond to three-, four- and five-loop running.}
\end{figure}

Fig.~\ref{fig::mudecbc} shows the dependence of $\alpha^{(5)}_s(M_Z)$ on the
decoupling scales for the charm (left) and bottom (right) quark mass using
three-, four- and five-loop running. It is easily obtained with the help
  of \verb|As5MZ| introduced above. In the case of the bottom quark one
observes an almost $\mu^{(b)}_{\rm dec}$-independent behaviour at five loops. In fact, the
variation of $\alpha^{(5)}_s(M_Z)$ in the considered range of $\mu^{(b)}_{\rm dec}$ is below
0.03\%.  Also in the case of the charm quark the curves become
flatter when including higher order quantum correction. However, at the
five-loop order the variation of $\alpha^{(5)}_s(M_Z)$ still amounts to
about 0.3\% which corresponds to $\delta_c=0.00030$ as mentioned above.


\subsubsection{$\Lambda$ and $\alpha_s$}

{\tt RunDec} and {\tt CRunDec} implement two different routines which allow to obtain
the QCD scale parameter $\Lambda^{(n_f)}$ for a given
$\alpha_s^{(n_f)}(\mu)$. The numerical results are slightly different
since one of the routines uses an explicit solution of the form
$\Lambda^{(n_f)} = f(\alpha_s^{(n_f)})$ whereas the other
looks for an implicit solution of $\alpha_s^{(n_f)} = f(\Lambda^{(n_f)})$.
Using

\begin{verbatim}
Do[ expl[i] = LamExpl[asMz /. NumDef, Mz /. NumDef, 5, i];
    impl[i] = LamImpl[asMz /. NumDef, Mz /. NumDef, 5, i];
    diff[i] = Abs[expl - impl]
    ,{i,1,5} ]
\end{verbatim}
we obtain the values shown in Tab.~\ref{tbl:lam_loops}. 
As expected, the difference decreases with increasing number of
loops.\footnote{The one-loop relation trivially leads to identical
  results for {\tt LamExpl} and {\tt LamImpl}.}

\begin{table}[t]
  \begin{center}
    \begin{tabular}{c | r r r}
      $\#$ loops & \verb|LamExpl| & \verb|LamImpl| & difference \\
      \hline
      1 & $88.35$ MeV & $88.35$ MeV & $0$ MeV \\
      2 & $209.86$ MeV & $227.51$ MeV & $17.65$ MeV \\
      3 & $210.10$ MeV & $209.54$ MeV & $0.56$ MeV \\
      4 & $209.78$ MeV & $209.53$ MeV & $0.25$ MeV \\
      5 & $209.80$ MeV & $209.87$ MeV & $0.07$ MeV \\
    \end{tabular}
    \caption{\label{tbl:lam_loops}$\Lambda^{(5)}$ calculated from
      $\alpha_s^{(5)}(M_Z)$ using both the explicit and implicit routine. The
      difference decreases with increasing number of loops.}
  \end{center}
\end{table}

In the next example we use the PDG value for the strong coupling constant (see
Eq.~(\ref{eq::input})) and we calculate $\Lambda^{(n_f)}$ for $n_f = 3,4,5,6$.
To do so we provide the following routine, which requires as input the strong
coupling constant and its uncertainty for $n_f$ active flavours at the scale
$\mu$. Furthermore the number of loops has to be specified.

\begin{verbatim}
CalcLam[as_,aserr_,mu_,nf_,nloops_] := Module[{expl4, impl4, expl, impl,
        explplus, implplus, explminus, implminus, mean, mean4,
        truncuncert, diffuncert, expuncert},
    expl  = LamExpl[as, mu, nf, nloops];
    impl  = LamImpl[as, mu, nf, nloops];
    expl4 = LamExpl[as, mu, nf, nloops-1];
    impl4 = LamImpl[as, mu, nf, nloops-1];
    explplus  = Abs[LamExpl[as + aserr, mu, nf, nloops] - expl];
    implplus  = Abs[LamImpl[as + aserr, mu, nf, nloops] - impl];
    explminus = Abs[LamExpl[as - aserr, mu, nf, nloops] - expl];
    implminus = Abs[LamImpl[as - aserr, mu, nf, nloops] - impl];
    mean  = (expl + impl)/2;
    mean4 = (expl4 + impl4)/2;
    truncuncert = Abs[mean - mean4];
    diffuncert  = Abs[expl - impl]/2;
    expuncert   = Max[explplus, implplus, explminus, implminus];
    Return[{mean, expuncert, truncuncert, diffuncert}];
];
\end{verbatim}
The uncertainties introduced through truncation of the perturbative series is
taken to be the difference between $n$- and $(n-1)$-loop results. We use both
\verb|LamExpl| and \verb|LamImpl| and take the mean value as the result and
assign an uncertainty due to half of the difference of both methods.  For the
uncertainty induced by $\delta\alpha_s$ we cite the larger one obtained by the two
methods. For $n_f = 5$ we obtain $\Lambda^{(5)}$ from

\begin{verbatim}
lam5 = CalcLam[asMz /. NumDef, 0.0011, Mz /. NumDef, 5, 5];
\end{verbatim}
For the other values of $n_f$ the code can be found in the supplementary
files. Let us only mention that, in order to obtain $\alpha_s^{(n_f)}(\mu)$ we
perform a decoupling of the top, bottom and charm quarks at the scales $M_t$,
$m_b(m_b)$ and $3$~GeV.  The results for $\Lambda^{(n_f)}$ and the
corresponding uncertainties can be found in Tab.~\ref{tbl:lam_nf}.  Our
central values and uncertainties for $n_f=4,5,6$ are identical to the ones
from PDG~\cite{Olive:2016xmw}. For $n_f = 3$ a minor difference is observed
which is due to the different decoupling scale chosen for the charm quark.  In
Ref.~\cite{Olive:2016xmw} $\mu^{(c)}_{\rm dec}=1.3$~GeV has been used which
results in $\Lambda = 332\pm17$~MeV. We can reproduce this result using our
code with four-loop accuracy.  Since $\alpha_s^{(3)}(\mu^{(c)}_{\rm dec}=1.3~\mbox{GeV})$
is already quite large we prefer to decouple the
charm quark for $\mu^{(c)}_{\rm dec}=3$~GeV.

\begin{table}[t]
  \begin{center}
    \begin{tabular}{c | r r r r }
      $n_f$ & $\Lambda^{(n_f)}$ & exp. uncertainty & $\delta_{\mathrm{trunc}}$
      & $\delta_{\mathrm{diff}}$ \\
      \hline
      3 & $336$ MeV & $17$ MeV & $1$ MeV & $1$ MeV \\
      4 & $292$ MeV & $16$ MeV & $2$ MeV & $1$ MeV \\
      5 & $210$ MeV & $13$ MeV & $0$ MeV & $0$ MeV \\
      6 & $89$ MeV & $6$ MeV & $0$ MeV & $0$ MeV \\
    \end{tabular}
    \caption{\label{tbl:lam_nf}$\Lambda^{(n_f)}$ calculated from
      $\alpha_s^{(5)}(M_Z)=0.1181 \pm 0.0011$. The uncertainties due
      to truncation of the perturbative series and due to the two different
      methods are shown. For our analysis we use five-loop accuracy.}
  \end{center}
\end{table}

In a recent publication the ALPHA collaboration has presented results for $\Lambda^{(3)}$ and
$\alpha_s^{(5)}(M_Z)$~\cite{Bruno:2017lta}.  In the following example we 
use $\Lambda^{(3)}= (332 \pm 14)$~MeV~\cite{Bruno:2017lta}, 
compute $\alpha_s^{(5)}(M_Z)$ using two approaches, and compare to
Ref.~\cite{Bruno:2017lta}.
In the first one we compute $\Lambda^{(4)}$ and
$\Lambda^{(5)}$ and finally extract $\alpha_s^{(5)}(M_Z)$.
The corresponding {\tt RunDec} commands read

\begin{verbatim}
Lambda3 = 0.332;
nloops  = 5;
Lambda4 = DecLambdaUp[Lambda3, muc /. NumDef, 3, nloops];
Lambda5 = DecLambdaUp[Lambda4, mub /. NumDef, 4, nloops];
aslam   = AlphasLam[Lambda5, Mz /. NumDef, 5, nloops];
\end{verbatim}
We obtain $\alpha_s^{(5)}(M_Z) = 0.1180$ in excellent
agreement with 
\begin{eqnarray}
  \alpha_s^{(5)}(M_Z) = 0.1179 \pm 0.0010 \pm 0.0002
  \label{eq::asMzlattice}
\end{eqnarray}
from Ref.~\cite{Bruno:2017lta}. We can also reproduce the uncertainties
which origin from the one of $\Lambda^{(3)}$ and the use of perturbation
theory (see the {\tt Mathematica} and {\tt C++} code in the ancillary
files.).

In the second approach one can use \verb|AlphasLam| to obtain $\alpha_s^{(3)}$
and calculate $\alpha_s^{(5)}(M_Z)$ by running and decoupling as described in
Section~\ref{sub::tau}.  Using five-loop accuracy, we obtain
\begin{eqnarray}
  \alpha_s^{(5)}(M_Z) = 0.1177 \pm 0.0009 \pm 0.0002 \pm 0.0003
  \,,
  \label{eq::asMzrundec}
\end{eqnarray}
where the last two uncertainties are due to truncation of the perturbative
series and the scale uncertainty. Both the central value and the three
uncertainties are obtained with the help of

\begin{verbatim}
Lambda3      = 0.332;
Lambda3plus  = Lambda3 + 0.014;
Lambda3minus = Lambda3 - 0.014;
mudecc       = 3.0;
mudecb       = mub /. NumDef;
nloops       = 5;
f = 3;

As3Lambda[lamin_?NumberQ, scale1_?NumberQ, scale2_?NumberQ, n_?IntegerQ] 
    := Module[{as31, as32},
    as31 = AlphasLam[lamin, scale1, 3, n];
    as32 = AlphasExact[as31, scale1, scale2, 3, n];
    Return[as32];
];

As5MZfromLambda[lamin_?NumberQ, scale1_?NumberQ, thr1_?NumberQ,
                thr2_?NumberQ, n_?IntegerQ] := Module[{as3, as5},
    as3 = As3Lambda[lamin, scale1, thr1, n];
    as5 = As5MZ[as3, thr1, thr1, thr2, n];
    Return[as5];
];

as5         = As5MZfromLambda[Lambda3, mudecc, mudecc, mudecb, nloops];
expuncert = Max[
              Abs[as5 - As5MZfromLambda[Lambda3plus, mudecc, mudecc, mudecb, nloops]],
              Abs[as5 - As5MZfromLambda[Lambda3minus, mudecc, mudecc, mudecb, nloops]] ];
truncuncert = Abs[as5 - As5MZfromLambda[Lambda3, mudecc, mudecc, mudecb, nloops-1]];
step = 1;
tmp = Map[ As5MZfromLambda[Lambda3, #, mudecc, mudecb, nloops]&,
           Range[mudecc/f, f*mudecc, step]];
scaleuncertlam = Max[ tmp ] - Min[ tmp ];
tmp = Map[ As5MZfromLambda[Lambda3, mudecc, #, mudecb, nloops]&, 
           Range[mudecc/f, f*mudecc, step]];
scaleuncertc = Max[ tmp ] - Min[ tmp ];
tmp = Map[ As5MZfromLambda[Lambda3, mudecc, mudecc, #, nloops]&, 
           Range[mudecb/f, f*mudecb, step]];
scaleuncertb = Max[ tmp ] - Min[ tmp ];
totalscaleuncert = Sqrt[scaleuncertc^2 + scaleuncertb^2 + scaleuncertlam^2];
\end{verbatim}
We have introduced \verb|as5MZfromLambda| which calculates
$\alpha_s^{(3)}(\mu)$. It is based on \verb|as5MZ|, which has been introduced
in Section~\ref{sub::tau} to compute $\alpha_s^{(5)}(M_Z)$ from
$\alpha_s^{(3)}(m_{\tau})$. Note that 
the central values in Eqs.~(\ref{eq::asMzlattice}) and~(\ref{eq::asMzrundec})
agree within the second uncertainty of Eq.~(\ref{eq::asMzlattice}) which is due
to the use of perturbation theory when relating $\Lambda^{(3)}$ to
$\Lambda^{(5)}$.  Note, however, that the corresponding uncertainty in our
second method, which is the one we recommend for such calculations, amounts to
$\sqrt{0.0002^2+0.0003^2}\approx 0.0004$ and is twice as large.

As a last example in this subsection we use $\Lambda^{(3)}$ from
Ref.~\cite{Bruno:2017lta} and compute $\alpha_s^{(3)}(m_{\tau})$
with the help of the following commands

\begin{verbatim}
astau       = AlphasLam[Lambda3, Mtau /. NumDef, 3, 5];
astau4      = AlphasLam[Lambda3, Mtau /. NumDef, 3, 4];
astauplus   = Abs[AlphasLam[Lambda3plus, Mtau /. NumDef, 3, 5] - astau];
astauminus  = Abs[AlphasLam[Lambda3minus, Mtau /. NumDef, 3, 5] - astau];
truncuncert = Abs[astau - astau4];
expuncert   = Max[astauplus, astauminus];
\end{verbatim}
We obtain $\alpha_s^{(3)}(m_\tau) = 0.3119 \pm 0.0073 \pm 0.0040$, where the
first uncertainty is due to the one of $\Lambda^{(3)}$
and the second one due to the truncation of the perturbative series.
For comparison, we also show the value obtained from
computing in a first step $\alpha_s^{(3)}(3~\mbox{GeV})$
and the subsequent running from 3~GeV to $m_\tau$. We obtain
$\alpha_s^{(3)}(m_\tau) = 0.3125 \pm 0.0074 \pm 0.0019$.
The results are consistent with each other. However, the uncertainties
from the truncation of the perturbative series differ by a factor two
which indicates that $\alpha_s$ is relatively large. It seems to be
advantageous to use the relation between $\Lambda$ and $\alpha_s$
at higher scales and use the renormalization group running to arrive at
$\mu=m_\tau$. 


\subsubsection{Running and decoupling for $m_b$}

For the decay of the Standard Model Higgs boson to bottom quarks it is
necessary to evaluate the $\overline{\rm MS}$ bottom quark mass
at an energy scale which is of the order of $M_H$.
The uncertainty of the decay rate depends crucially on the accuracy
of $m_b$ and thus it is important to consistently propagate the uncertainties
from low to high energies. In Ref.~\cite{Chetyrkin:2009fv} one finds the
following result for the bottom quark mass
\begin{eqnarray}
  m_b^{(5)}(10~\mbox{GeV})=\left(3610-{\alpha_s-0.1189\over0.002}\cdot12\pm11\right)~\mbox{MeV}
  \,,
\end{eqnarray}
which translates to

\begin{eqnarray}
  m_b^{(5)}(M_H)=2771\pm\left.8\right|_{m_b}\pm\left.15\right|_{\alpha_s}~\mbox{MeV}
  \,.
  \label{eq::mbMH}
\end{eqnarray}
The corresponding {\tt Mathematica} commands to obtain the central
value are given by

\begin{verbatim}
  mb10 = 3.610 - 12/1000*((asMz /. NumDef) - 0.1189)/0.002;
  mu10 = 10;
  nloops = 5;
  as10 = AlphasExact[asMz /. NumDef, Mz /. NumDef, mu10, 5, nloops];
  asMh = AlphasExact[asMz /. NumDef, Mz /. NumDef, Mh /. NumDef, 5, nloops];
  mMS2mMS[mb10, as10, asMh, 5, nloops]
\end{verbatim}
The code which calculates the uncertainties and the {\tt C++} code can be
found in the supplementary files. 

As a further example let us consider the bottom quark mass
at the scale $\mu=M_t$ in the six-flavour theory. With the
help of

\begin{verbatim}
  mL2mH[mb10, as10, mu10, {{6, Mt /. NumDef, 2*Mt /. NumDef}}, Mt /. NumDef, 5]
\end{verbatim}
one obtains $m_b^{(6)}(M_t) = 2670$~MeV.


\subsection{Quark mass relations}


\subsubsection{Relation between threshold and $\overline{\rm MS}$ quark masses}

In this subsection we provide {\tt RunDec} (and {\tt CRunDec}) commands which
allow to check the numbers in Tab.~3 of Ref.~\cite{Marquard:2016dcn}.  To do so
we use the newly implemented routines for the conversion between threshold and
$\overline{\rm MS}$ quark masses.  In addition we allow for an uncertainty in the
threshold mass and $\alpha_s^{(5)}(M_Z)$ and compute the resulting
uncertainty of the $\overline{\rm MS}$ quark mass in the output.
The {\tt RunDec} commands which fulfill this task are given by

\begin{verbatim}
xerr = 0.002;
mbSI = mub /. NumDef;
mtOS = Mt /. NumDef;
mcSI = muc /. NumDef;

mThr2mSI[mThr_,merr_,asmz_,aserr_,nl_,nloop_,scheme_] := Module[
    {as, asp, asm, muf, mSI, mcentral, msoserr, expuncert, muncert, asuncert},
    as[mu_] := Switch[nl,
                5, AlphasExact[asmz, Mz /. NumDef, mu, 5, 4],
                4, AlphasExact[DecAsDownSI[AlphasExact[asmz, Mz /. NumDef, 2*mbSI, 5, 4],
                   mbSI, 2*mbSI, 4, 4], 2*mbSI, mu, 4, 4],
                3, AlphasExact[DecAsDownSI[AlphasExact[DecAsDownSI[AlphasExact[
                    asmz, Mz /. NumDef, 2*mbSI, 5, 4], mbSI, 2*mbSI, 4, 4],
                   2*mbSI, 3.0, 4, 4], mcSI, 3.0, 3, 4], 3.0, mu, 3, 4]
                ];
    asp[mu_] := Switch[nl,
                5, AlphasExact[asmz + aserr, Mz /. NumDef, mu, 5, 4],
                4, AlphasExact[DecAsDownSI[AlphasExact[asmz + aserr, Mz /. NumDef,
                   2*mbSI, 5, 4], mbSI, 2*mbSI, 4, 4], 2*mbSI, mu, 4, 4],
                3, AlphasExact[DecAsDownSI[AlphasExact[DecAsDownSI[AlphasExact[
                    asmz + aserr, Mz /. NumDef, 2*mbSI, 5, 4], mbSI, 2*mbSI, 4, 4],
                    2*mbSI, 3.0, 4, 4], mcSI, 3.0, 3, 4], 3.0, mu, 3, 4]
                ];
    asm[mu_] := Switch[nl,
                5, AlphasExact[asmz - aserr, Mz /. NumDef, mu, 5, 4],
                4, AlphasExact[DecAsDownSI[AlphasExact[asmz - aserr, Mz /. NumDef,
                   2*mbSI, 5, 4], mbSI, 2*mbSI, 4, 4], 2*mbSI, mu, 4, 4],
                3, AlphasExact[DecAsDownSI[AlphasExact[DecAsDownSI[AlphasExact[
                    asmz - aserr, Mz /. NumDef, 2*mbSI, 5, 4], mbSI, 2*mbSI, 4, 4],
                    2*mbSI, 3.0, 4, 4], mcSI, 3.0, 3, 4], 3.0, mu, 3, 4]
                ];
    muf := Switch[nl, 5, 80.0, 4, 2.0, 3, 2.0];
    
    mSI[m_,asnl_,err_] := Switch[scheme,
                "1S", m1S2mSI[m, {}, asnl, nl, nloop, 1+err],
                "PS", mPS2mSI[m, {}, asnl, muf, nl, nloop, 1+err],
                "RS", mRS2mSI[m, {}, asnl, muf, nl, nloop, 1+err],
                "RSp", mRSp2mSI[m, {}, asnl, muf, nl, nloop, 1+err]];


    mcentral := mSI[mThr, as, 0];
    msoserr := Abs[mSI[mThr, as, xerr] - mcentral];
    If[merr =!= 0,
        muncertplus := Abs[mcentral - mSI[mThr + merr[[1]], as, 0]],
        muncertplus = 0;
    ];
    If[Length[merr] > 1,
        muncertminus := Abs[mcentral - mSI[mThr - merr[[2]], as, 0]],
        muncertminus = muncertplus;
    ];
    If[aserr =!= 0,
        asuncert := Max[ Abs[mcentral - mSI[mThr, asp, 0]], 
                         Abs[mcentral - mSI[mThr, asm, 0]] ],
        asuncert = 0;
    ];
    Return[{mcentral, muncertplus, muncertminus, asuncert, msoserr}];
];
\end{verbatim}
Using this routine we can compute the Tab.3a of~\cite{Marquard:2016dcn} with
the help of

\begin{verbatim}
Do[ mfromPS = First[mThr2mSI[168.049, 0, asMz /. NumDef, 0, 5, i, "PS"]];
    mfrom1S = First[mThr2mSI[172.060, 0, asMz /. NumDef, 0, 5, i, "1S"]];
    mfromRS = First[mThr2mSI[166.290, 0, asMz /. NumDef, 0, 5, i, "RS"]];
    mfromRSp = First[mThr2mSI[171.785, 0, asMz /. NumDef, 0, 5, i, "RSp"]];
    Print[i, "       ", mfromPS, "  ", mfrom1S, "  ", mfromRS, "  ", mfromRSp];
    ,{i,1,4} ];
\end{verbatim}
The remaining three tables are obtained with similar code (see ancillary files).

As a further example let us assume that the top quark PS mass has
been extracted from experimental data (e.g. from the threshold cross section
of a future electron positron linear collider) to 
\begin{eqnarray}
  m_t^{\rm PS}&=&168.049 \pm 0.100~\mbox{GeV}
  \,.
\end{eqnarray}
Using $\alpha_s^{(5)}(M_Z)$ from Eq.~(\ref{eq::input}) and
\verb|mThr2mSI| leads to
\begin{eqnarray}
  m_t^{\overline{\rm MS}}&=&163.508 \pm 0.095_{\delta m_t^{\rm PS}} 
  \pm 0.043_{\delta\alpha_s}~\mbox{GeV}
  \,.
\end{eqnarray}
It is interesting to note that an uncertainty of 0.0011 in
$\alpha_s^{(5)}(M_Z)$ induces an uncertainty of about 40~MeV
into the $\overline{\rm MS}$ top quark mass.

In Ref.~\cite{Beneke:2014pta,Beneke:2016oox} the PS bottom quark mass
has been determined to  $m_b^{\rm PS} = 4.532^{+0.013}_{-0.039}$~GeV
using QCD sum rules. With the help of

\begin{verbatim}
mbPS         = 4.532;
delmbPSplus  = 0.013;
delmbPSminus = 0.039;
as           = asMz /. NumDef;
alsuncert    = 0.0013;
mThr2mSI[mbPS, {mbPSplus, mbPSminus}, asMz /. NumDef, alsuncert, 4, 4, "PS"];
\end{verbatim}
we obtain $m_b(m_b) = 4.209^{+0.014}_{-0.036}$~GeV,
in good agreement with the result given in~\cite{Beneke:2016oox}
$m_b(m_b) = 4.203^{+0.016}_{-0.034}$~GeV.


\subsubsection{$\overline{\rm MS}$-OS relations including light quark mass effects}

Light quark mass effects in the $\overline{\rm MS}$-OS relation of the top and
bottom quark masses can be computed with the help of the following commands

\begin{verbatim}
as6[mu_] := AlL2AlH[asMz /. NumDef, Mz /. NumDef, 
                    {{6, Mt /. NumDef, 2*Mt /. NumDef}}, mu, 5];
as5[mu_] := AlphasExact[asMz /. NumDef, Mz /. NumDef, mu, 5, 5];

mut  = 163.0;
Mt4  = mMS2mOS[mut, {}, as6[mut]*XXX, mut, 6, 4];
Mtb  = mMS2mOS[mut, {mub /. NumDef}, as6[mut]*XXX, mut, 6, 4];
Mtbc = mMS2mOS[mut, {{mub /. NumDef, mub /. NumDef}, {mc3 /. NumDef, 3.0}},
               as6[mut]*XXX, mut, 6, 4];
delb = Expand[(Mtb - Mt4)];
delc = Expand[(Mtbc - Mtb)];

Mb4  = mMS2mOS[mub /. NumDef, {}, as5[mub /. NumDef]*XXX, mub /. NumDef, 5, 4];
Mbc  = mMS2mOS[mub /. NumDef, {{mc3 /. NumDef,3.0}}, 
               as5[mub /. NumDef]*XXX, mub /. NumDef, 5, 4];
delc = Expand[(Mbc - Mb4)];
\end{verbatim}
The label \verb|XXX| has been introduced to separate the different loop
orders and the mass effects are contained in the quantities \verb|delc| and
\verb|delb|. For the top and bottom quark mass we have
\begin{eqnarray}
  M_t &=& (163 + 7.509_{1l} + 1.603_{2l} + 0.495_{3l} + 0.195_{4l}\nonumber\\
  &&\mbox{} + 0.008_{2lb} + 0.002_{2lc} + 0.010_{3lb} +
  0.004_{3lc})~\mathrm{GeV}
  \,,\nonumber\\
  M_b &=& (4.163 + 0.398_{1l} + 0.206_{2l} + 0.160_{3l} + 0.136_{4l}
  + 0.007_{2lc} + 0.015_{3lc})~\mathrm{GeV}
  \,,
\end{eqnarray}
where the subscripts are self-explanatory.
In all cases the mass corrections at three loops are larger than
at two loops which suggests that there is no convergence in
the mass correction terms. The situation is improved after
decoupling the heavy quarks and using $\alpha_s^{(3)}$ as an expansion
parameter (see, e.g., Ref.~\cite{Ayala:2014yxa}).



\section{\label{sec::sum}Summary}

This article describes the improvements implemented in versions~3 of the
programs {\tt RunDec} and {\tt CRunDec}.  This concerns in particular the
five-loop corrections to the running of strong coupling constant and the quark
masses, the four-loop correction to the decoupling of heavy quarks from the
running, the four-loop corrections to the relation between the $\overline{\rm
  MS}$ and on-shell heavy quark mass, and various relations between the
$\overline{\rm MS}$ and on-shell heavy quark mass to so-called threshold
masses. Furthermore, light quark mass effects are implemented in the
$\overline{\rm MS}$-on-shell relation to three loops.  We provide a large
number of examples which exemplify the use of both the {\tt Mathematica} and
{\tt C++} version. It is straightforward to apply modifications and include
them in other codes.


\section*{Acknowledgements}

We thank Konstantin Chetyrkin for providing analytic results for
the five-loop QCD beta function and quark anomalous dimension.
This work is supported by the BMBF through Grant No. 05H15VKCCA.





\end{document}